\documentstyle[mprocl,psfig]{article}

\def\Journal#1#2#3#4{{#1} {\bf #2}, #3 (#4)}

\def\NPB{{\em Nucl. Phys.} B}
\def\PLB{{\em Phys. Lett.}  B}

\def\PRD{{\em Phys. Rev.} D}

\def\be{\begin{equation}}
\def\ee{\end{equation}}
\def\bea{\begin{eqnarray}}
\def\eea{\end{eqnarray}}

\def\chiral{{\bf\rm C}}
\def\wilson{{\bf\rm W}}
\def\dirac{{\bf\rm D}}
\def\ham{{\bf\rm H}}
\def\mbham{{\cal H}}
\def\wbrU{{\rangle^{\rm WB}_U}}
\def\wbrUg{{\rangle^{\rm WB}_{U^g}}}
\def\wbr1{{\rangle^{\rm WB}_1}}
\def\wbl1{{{}^{\rm WB}_{\ \ 1}\!\langle}}

\begin{document}

\rightline{FSU-SCRI-97-84}

\title{Chiral Gauge Theories in the Overlap Formalism}

\author{ Rajamani Narayanan }

\address{Supercomputer Computations Research Institute\\
The Florida State University \\
Tallahassee, FL 32306, USA\\E-mail: rnaray@scri.fsu.edu}

\maketitle
\abstracts{ 
The overlap formula for the chiral determinant is presented and the
realization of gauge anomalies and gauge field toplogy in this context
is discussed. The ability of the overlap formalism to deal with
supersymmetric theories and Majorana-Weyl fermions is outlined. Two
applications of the overlap formalism are discussed in some detail.
One application is the computation of a fermion number violating
process in a two dimensional U(1) chiral gauge theory. The second
application is a measurement of the probability ditribution of the
index of the chiral Dirac operator in four dimensional pure SU(2)
lattice gauge theory.}

\section{Introduction}

Significant progress has been made in the lattice formalism
of chiral gauge theories using the overlap formalism~\cite{overlap}.
Using this formalism, a fermion number violating process has
been computed on the lattice in a two dimensional abelian chiral
model~\cite{chiral1,chiral2}, thereby demonstrating 
the validity of the overlap formula. The overlap formalism was
inspired by two independent articles~\cite{Kaplan,Slavnov}.
In the first article~\cite{Kaplan}, 
a $d+1$ dimensional Dirac fermion with a mass
term incorporating a defect is used to realize a localized chiral
fermion in $d$ dimensions. This idea of localization orginally appears
in a paper by Callan and Harvey~\cite{Callan} and 
was later used for realizing chiral fermions in the
context of condensed matter physics~\cite{Fradkin}.
The second article~\cite{Slavnov}
 provides a Pauli-Villars regularization for the
fermion determinant in perturbation theory
when the fermion is in the sixteen dimensional
Weyl representation of the SO(10) gauge group. 
There is a need for an infinite number of Pauli-Villars fields to
regulate the theory, reminiscent of the need for one extra dimension
in the first article~\cite{Kaplan} to realize a localized chiral fermion.

The need to have an infinite number of degrees of freedom per space time
point (one extra dimension in the context of the first article~\cite{Kaplan}
 and an
infinite number of Pauli-Villars fields in the context of the second
article~\cite{Slavnov}) is not surprising for two closely related reasons: 
\begin{itemize}
\item Existence of gauge anomalies:
An extra infinity is needed to realize the gauge anomaly
present when the fermion is in a anomalous representation of the gauge
group. Without this extra infinity, the anomaly is forced to come out
as a consequence of the ultra-violet regularization of the theory even
though the anomaly is independent of the regularization scheme.
\item Fermion number violation: 
In a non-perturbative regularization such as the lattice formalism,
the number of space time points are finite and 
an infinite number of fields is needed to realize the non-trivial index of
the chiral Dirac operator due to the Atiyah-Singer index theorem.
The index is a property of the Dirac operator per gauge field configuration
and therefore has to be realized on a finite lattice. 
\end{itemize}

The central problem in the non-perturbative formalism of chiral gauge
theories is to write down a formula for the fermionic
determinant where the fermion is in some complex representation of the
gauge group. Let $\chiral(U)$ denote the chiral Dirac operator coupled
to a background gauge field $U$ in some complex representation.
Clearly $\chiral(U)$ does not have an eigenvalue problem since
it is a map between $(0,{1\over 2})$ and $({1\over 2},0)$ spinors
under the Lorentz group. Therefore ``det''$\chiral(U)$ cannot be
represented as a product of eigenvalues and one cannot
make use of the zeta function regulator. 
The existence of gauge anomalies is related to this particular nature
of the operator.
Because $\chiral(U)$ is a map between two spaces, the operator
can have a non-trivial index, i.e., the infinite matrix $\chiral(U)$
is sometimes square and sometimes rectangular and the difference
between the infinite number of columns and rows of this matrix is
a finite number that is dictated by the topology of the gauge field
background. This interesting property is responsible for fermion
number violation. Any formula for the chiral determinant has to
respect the above mentioned properties. 

\section{Overlap formalism}

In this section, the overlap formalism will be presented as a representation
of the chiral determinant in the continuum. Then a straightforward lattice
regularization of the overlap will be described. Gauge anomaly and gauge field
topology in the context of the overlap formalism will then follow. Extention
of the overlap formalism to the most basic fermionic object in $2+8k$ dimensions,
namely Majorana-Weyl fermions, will also be presented with the aim of extending
the overlap formalism to include supersymmetric theories.

\subsection{Chiral determinant as an overlap of two vacua}

In Euclidean space, ``det''$\chiral(U)$ is a complex functional of
$U$ and its phase in general is not gauge invariant and carries the
information about the anomaly associated with the particular
complex representation of the gauge group. Any formula for ``det''$\chiral(U)$
should reproduce this anomaly. In addition the ``det''$\chiral(U)$
should be zero for a large class of gauge fields where the matrix
$\chiral(U)$ is not square. Further the formula for left and right handed
chiral fermions in the same representation should be related by complex
conjugation. The formula should also be in accordance with the standard
discrete symmetries of parity and charge conjugation. 

The overlap formula
is one that satisfies all the above requirements~\cite{overlap}. In
the overlap formalism, ``det''$\chiral(U)$ is represented as an
overlap of two many body states composed of fermions. This representation
is arrived at as follows. Let ${v_j}$ be a basis for the
$(0,{1\over 2})$ spinors and let ${u_i}$ be a basis for the
$({1\over 2}, 0)$ spinors. Formally,
``det''$\chiral=\det_{ij} \langle u_i | \chiral | v_j \rangle = 
\det_{ij} \langle u_i | w_j \rangle$ with $|w_j\rangle = \chiral |v_j\rangle$.
The numbers of basis vectors ${v_j}$ and ${u_i}$ are the dimensions of the
two spaces that is mapped by $\chiral$. Each vector has many components
and the number of components is dependent on our specific choice of
embedding. Specifically, $\langle u_i | w_j \rangle = \sum_\alpha
u^*_{i\alpha} w_{j\alpha}$ where $\alpha$ labels the components.
The number of components could very well be bigger than the dimension of
the two spaces mapped by $\chiral$ in our choice of embedding and this
will indeed be the case. The first step in deriving the overlap formula
is to make the following operator identification. With every vector
$u_i$, we associate a single particle fermion creation operator 
$\hat u_i = u_{i\alpha} a^\dagger_\alpha$. $a^\dagger_\alpha$ are
cannonical fermion creation operators that obey the following
standard commutation relations: $\{ a^\dagger_\alpha, a^\dagger_\beta \} =0$;
$\{ a_\alpha, a_\beta \} =0$;
$\{ a_\alpha, a^\dagger_\beta \} =\delta_{\alpha\beta}$.
Having made this operator identification, we now define two many body
states as $|+\rangle = \prod _j \hat w_j |0\rangle$
and $|-\rangle = \prod _i \hat u_i |0\rangle$ where $|0\rangle$ is the
state that is annihilated by all the destruction operators $a_\alpha$.
Now it is straighforward algebra to show that
$\langle - | + \rangle = \det_{ij} \langle u_i | w_j \rangle$
and this is the overlap formula.

A few remarks are in order at this time. The number of particles
making up the two many body states are equal to the dimensions of the
two spaces mapped by $\chiral$. These two dimensions need not be
equal for the formula to hold. If they are not equal the number of bodies
in the two many body states are not the same and overlap is zero
as expected. The number of fermion creation operators need not be equal
to either of the two dimensions. It is a consequence of the
specific embedding and is equal to or larger than the larger of the
two spaces.

Having represented the chiral determinant as an overlap of two many body
states, a procedure to obtain the many body states as ground states of
two auxiliary hamiltonians is now described. Let 
\begin{equation}
\mbham^\pm = a^\dagger \ham^\pm a
\label{eq:mham}
\end{equation}
be two many body Hamiltonians each describing a set of identical non-interacting
fermions. The state $|-\rangle=\prod_i \hat u_i | 0\rangle $ is simply
a choice of the basis for the $({1\over 2},0)$ spinors and this can
be obtained as a ground state of $\mbham^-$
by choosing
\begin{equation}
\ham^- = \pmatrix{ -1  & 0 \cr 0 & 1 } 
\label{eq:hamr}
\end{equation}
To obtain $|+\rangle=\prod_j \hat w_j | 0\rangle = \prod_j \chiral \hat v_j | 0\rangle$
as a ground state of $\mbham^+$,
a choice for the other single particle
Hamiltonian is
\begin{equation}
\ham^+ = \pmatrix{ m & \chiral \cr \chiral^\dagger &  -m} 
\label{eq:ham}
\end{equation}
with $m > 0$. 
Then,
\begin{equation}
\det\chiral \Leftrightarrow  \langle - | + \rangle
\label{eq:over} 
\end{equation}
The $\Leftrightarrow$ above is indicative of several points. The above formula is
only formal and will become the defintion of the chiral determinant when the
right hand side is properly regulated. The above formula is valid in the limit 
$m\rightarrow\infty$ and the parameter $m$ is to be thought of as an ultraviolet
regulator. Finally there is a irrelevant gauge field independent normalization
that depends on $m$ in the formula and as such the above formula is only valid
for ratios of determinants in different gauge fields backgrounds.

Eq.~\ref{eq:over} can be shown by first
noting that the ground states are obtained by filling 
the single particle negative energy eigenstates given by 
\begin{equation}
\ham^\pm \psi_i^{\pm} =
\lambda^{\pm}_i \psi_i^{\pm};\ \ \ \ \lambda^{\pm}_i < 0 
\label{eq:eigen}
\end{equation}
Clearly, $\lambda^-_i=-1$ for all $i$ and the eigenvectors with these eigenvalues
are
\begin{equation}
\psi^-_i = \pmatrix{ u_i \cr 0 \cr}
\label{eq:refstate}
\end{equation}
with the set $\{u_i\}$ forming a unitary matrix. 
To get $\lambda^+_j$ one needs to solve 
\begin{equation}
\chiral^\dagger\chiral v_j = ([\lambda^+_j]^2 - m^2 ) v_j
\end{equation}
In terms of these eigenfunctions, $\psi^+_j$ are given by
\begin{equation}
\psi_j^+  =
{1\over {\sqrt{N_{jj}}}} \pmatrix {
                    {1\over
                             {  \sqrt{\chiral\chiral^\dagger +m^2 }+m}
                    }\chiral v_j \cr 
					-v_j 
               };\ \ \ \ 
N_{jk} =  \langle v_j 
                \left [ 
                  {
                   {\chiral^\dagger \chiral}\over
                       {  (\sqrt {\chiral^\dagger \chiral +m^2 } 
               +m )^2}  
                  }  +1 
               \right ] v_k \rangle 
\end{equation}
with $N$ being a diagonal matrix. 
The overlap,
\begin{equation}
\langle -|+\rangle   = \det_{ij} ( \langle \psi_i^{-}| \psi_j^{+} \rangle )
 ={1\over {\sqrt{ \det N}}} ~\det_{ij} 
\langle u_i| {1\over{\sqrt{\chiral\chiral^\dagger +m^2}+|m|}}\chiral |v_j
\rangle 
\end{equation}
where  the first equality in the above equation is an identity resulting from
the canonical commutation relations of the fermion creation and destruction
operators~\cite{vacua}. 
Therefore, we have a 
formula for the chiral determinant as an overlap of
two vacua in Eq.~\ref{eq:over}. One of the vacua, namely $|-\rangle$ is 
simply a fixed reference state and the other one carries all the information
about the gauge field background. The formula uses the embedding of
the chiral Dirac operator in a vector like operator, namely $\ham^+$,
and achieves the desired goal. Regularization of the righ hand side of
Eq.~\ref{eq:over} amounts to a regularization of $\ham^+$ which can
be done in a straighforward manner since it is vector like. 

\subsection{Lattice regularization of the overlap}

The overlap formula in Eq.~\ref{eq:over} is a proper
definition for the chiral determinant 
only after regularization and a definition of the phase of the
two many body states. The single particle Hamiltonians in Eq.~\ref{eq:ham}
is equal to $\gamma_5 (\dirac + m)$ where $\dirac= \gamma_\mu(\partial_\mu
+ i A_\mu)$ is the massless Dirac operator in Euclidean space.
On the lattice, the massless Dirac operator is written as
\begin{equation}
\dirac(x\alpha i; y \beta j) = \sum_\mu \gamma^{\alpha\beta}_\mu {1\over 2} 
\Bigl[ \delta_{y,x+\hat\mu} U^{ij}_\mu(x) -
\delta_{x,y+\hat\mu} (U^\dagger_\mu(y))^{ij} \Bigr]
\label{eq:dirac}
\end{equation}
where $x,y$ are sites on the lattice, $\alpha,\beta$ are spin indices, 
$i,j$ are color indices and $U_\mu(x)$ is the parallel transporter along the
direction $\mu$ from $x$ to $x+\hat\mu$. In the free theory, $U=1$, this
operator is diagonal in momentum space and is given by $i\gamma_\mu\sin p_\mu$.
As is well known, this operator has many unwanted particles arising from the
zeros at the edge of the Brillouin zone $p_\mu=\pi$ and these unwanted particles
have to be removed. To do this, note that the overlap yields the determinant of
$\chiral$ only if the  Hamiltonian in Eq.~\ref{eq:ham} has $m>0$. 
It is clear that the overlap will be close to unity if $m$ was a large negative number
since $\ham^+$ approaches $\ham^-$. 
The unwanted particles 
can be removed by making the mass term momentum dependent in such a way that it is
positive when $p_\mu=0$ and is negative
when one or many of the $p_\mu=\pi$. This is achieved by replacing the
mass term by the usual Wilson term,
\begin{equation}
\begin{array}{rcl}
 m  & \rightarrow & (m - \wilson)(x\alpha i,y\beta j)\\
&&  = 
m \delta_{\alpha\beta} \delta_{ij} \delta_{xy} -
{1\over 2} \delta_{\alpha\beta}
\sum_\mu \Bigl[ 2\delta_{xy}\delta_{ij}-
\delta_{y,x+\hat\mu} U^{ij}_\mu(x) -
\delta_{x,y+\hat\mu} (U^\dagger_\mu(y))^{ij} \Bigr] \\
\label{eq:wilson}
\end{array}
 \end{equation}
In the free case, $U=1$, the mass term is replaced by a momentum dependent mass term
given by $ m - 2\sum_\mu \sin^2 {p_\mu\over 2} $. If we pick $m$ in the range
$0 < m < 2$, it is clear that the momentum dependent mass term is
positive when $p_\mu=0$ but is negative when one or more $p_\mu=\pi$. 
Hence the regulated Hamiltonian on the lattice is 
\begin{equation}
\ham^+ = \gamma_5 (\dirac - \wilson + m) ; \ \ \ \ 0 < m < 2
\label{eq:lham}
\end{equation}
with $\dirac$ and $\wilson$ given by Eq.~\ref{eq:dirac} and Eq.~\ref{eq:wilson}
respectively.  
In taking the continuum limit, $m$ should be kept fixed at some value in the
range $0 < m < 2$ so that it goes to infinity in physical units as the lattice
spacing is taken to zero. All values of $m$ in the range $0 < m < 2$ are expected
to yield the same continuum theory unless the theory being defined has some
marginally relevant parameters. The cutoff effects as one goes to the continuum
limit will depend on the actual value of $m$. 

Following arguments similar to the one for the chiral determinant above, it is
easy to show~\cite{overlap}
that the generating functional for fermions in an arbitrary gauge
background is given by
\begin{equation}
Z(\eta,\bar\eta) = \langle -| e^{\eta a^\dagger + \bar\eta a} |+\rangle 
\end{equation}
Therefore insertion of $a$ and $a^\dagger$ operators at appropriate places
inside the overlap result in correlation functions for fermions in a
gauge field background.

\subsection{Phase of the many body states and the gauge anomaly}

The regulated single particle Hamiltonian in Eq.~\ref{eq:lham} is a finite matrix
on a finite lattice and therefore $|+\rangle$ is a many body state with
a finite number of particles that depends upon the background gauge field.
To properly define the overlap it is necessary to define the phase of this
state. $|-\rangle$ is a reference state that does not depend on the gauge field
and its phase can be fixed once and for all by choosing the set $\{u_i\}$ to be
the identity matrix. A phase choice for $|+\rangle$
that is consistent with perturbation theory
is the Wigner-Brillouin phase choice. Under this phase choice, the overlap
$\wbl1 +|+ \wbrU$ is forced to be a real and positive quantity. 
Here $|+\wbr1$ is the free many body state whose phase is assumed fixed.
With respect to this state, the phase of $|+\wbrU$ for all $U$ is defined by
the Wigner-Brillouin convention. The subcript $U$ shows that the state depends
on the background gauge field and the superscript WB shows that it obeys
the Wigner-Brillouin convention. One can prove that this phase choice results
in the correct tranformation properties of the chiral determinant under parity
charge conjugation and global gauge transformations~\cite{overlap}.

Gauge anomalies now arise for a simple reason. Let $|+\wbrU$ and $|+\wbrUg$
be many body states with background gauge fields being $U$ and $U^g$ respectively
where $U^g$ is a gauge tranformation of $U$. Since the Hamiltonian in 
Eq.~\ref{eq:lham} undergoes a unitary rotation under this gauge tranformation,
the two states are simply related by a unitary rotation. That is
\begin{equation}
|+\wbrUg = {\cal G} |+\wbrU e^{i\phi(U;g)}
\end{equation}
where ${\cal G}$ is the unitary operator and the phase on the righthand side is
chosen so that $|+\wbrUg$ indeed obeys the  Wigner-Brillouin phase convention.
Now the overlap
\begin{equation}
\langle -|+\wbrUg = \langle -|+\wbrU e^{i\phi(U;g)} \prod_x g(x) 
\end{equation}
If $g(x)$ is not a global transformation, then the phase 
$e^{i\phi(U;g)} \prod_x g(x)$ is not unity showing that the chiral determinant
is not gauge invariant and the presence of a gauge anomaly. Basically what has
happened is the follows. The overlap with the Wigner-Brillouin phase convention
defines a proper functional of the gauge field background. The presence of
a gauge anomaly is the inability to make a proper functional that is also gauge
invariant. Anomaly and other perturbative quantities have been succesfully computed
in two dimensions and four dimensions using
the overlap~\cite{overlap,ictp}.

\subsection{Overlap and gauge field topology}

The specific nature of the operator $\chiral$ led us to a representation of
its determinant involving two many body states. As such we had to define
the phase of these states which resulted in a natural explanation of the
gauge anomaly. It is well known that the integration of the anomaly equation
results in the phenomenon of fermion number violation if the gauge field
carries non-trivial topology. Therefore it is natural to expect that topology
also arises from the fact that we are dealing with two many body states.
This is indeed the case. On a finite lattice, the single particle Hamiltonians
are finite matrices of size $2K\times 2K$ with $K=V\times N \times S$ where
$V$ is the volume of the lattice, $N$ is the size of the paticular 
representation of the gauge group and $S$ is the number of components of
a Weyl spinor (one in two dimensions and two in four dimensions). 
Then $|-\rangle$ is made up of $K$ particles. If $|+\wbrU$ is also
made up of $K$ particles, then the overlap is not zero in the generic case.
If the background gauge field is such that there are only $K-Q$ negative
energy states for $\ham^+$ then the overlap is zero. Any small perturbation
of the gauge field will not alter this situation. Further the overlap
$\langle -| a^\dagger_{i_1}\cdots a^\dagger_{i_Q} |+ \wbrU$ will not be
zero in the generic case if the fermion is in the
fundamental representation of the gauge group
showing that there is a violation of fermion
number by $Q$ units. Clearly this results in a classification of gauge
fields on a finite lattice into topological classes labelled by $Q$. 
The topological nature of the gauge fields using the overlap has been
investigated both in two and four dimensions~\cite{overlap,daemi}. 

\subsection{Supersymmetry and Majorana-Weyl fermions}

The overlap formalism enables one to write down a SU(N) gauge theory on the lattice
coupled to a single adjoint multiplet of left-handed Weyl fermions. The continuum limit
of this theory should be supersymmetric and no fine tuning is needed to achieve this
limit making it more attractive than a previous approach using Wilson fermions on
the lattice~\cite{Curci}. Using the overlap, one could go a step further and 
formulate a gauge theory coupled to Majorana-Weyl fermions, the most basic fermion in
$2+8k$ dimensions~\cite{MWeyl}. One application of this would be the ten dimensional
N=1 supersymmetric Yang Mills theory. Another application would be the investigation
of possible fermion bilinear condensates in two dimensional non-abelian gauge theories
coupled to Majorana-Weyl fermions. The underlying structure seems to a mod(2) index
associated with the Majorana-Weyl operator. 

The overlap formalism for Majorana-Weyl fermions can be obtained using 
the factorization of
the overlap for Weyl fermions. In the chiral basis,
\begin{equation}
\mbham^+ = \pmatrix{a_1^\dagger & a_2^\dagger \cr}
\pmatrix{m-\wilson & \chiral \cr \chiral^\dagger & \wilson-m \cr}
\pmatrix{a_1 \cr a_2 \cr}
\end{equation}with
\begin{equation}
\chiral(x\alpha i; y \beta j) = \sum_\mu \sigma^{\alpha\beta}_\mu {1\over 2} 
\Bigl[ \delta_{y,x+\hat\mu} U^{ij}_\mu(x) -
\delta_{x,y+\hat\mu} (U^\dagger_\mu(y))^{ij} \Bigr]
\end{equation}
If the Weyl fermion is in a real representation of the gauge group in $2+8k$ dimensions,
then the Weyl operator is skew-symmetric, i.e., $\chiral^t = -\chiral$. Further
$\wilson^t=\wilson$. Using the above relations, one can show that 
under the canonical transformation, $a_1={\xi-i\eta\over \sqrt{2}}$
and $a_2={\xi^\dagger-i\eta^\dagger\over \sqrt{2}}$, 
\begin{equation}
\begin{array}{rcl}
\mbham^+ = && {1\over 2} \pmatrix{\xi^\dagger & \xi \cr}
\pmatrix{m-\wilson & \chiral \cr \chiral^\dagger & \wilson-m \cr}
\pmatrix{\xi \cr \xi^\dagger \cr} \\ 
&& +
{1\over 2}\pmatrix{\eta^\dagger & \eta \cr}
\pmatrix{m-\wilson & \chiral \cr \chiral^\dagger & \wilson-m \cr}
\pmatrix{\eta \cr \eta^\dagger \cr}
\\
\end{array}
\label{eq:MWeyl}
\end{equation}
Therefore, the many body Hamiltonian factorizes into two identical pieces
each one corresponding to a single Majorana-Weyl fermion.

A mod(2) index naturally arises in the context of Majorana-Weyl fermion operator.
Clearly, the many body Hamiltonian for a single Majorana-Weyl fermion in
Eq.~\ref{eq:MWeyl} does not
conserve particle number. But it does conserve particle number, modulo two.
The ground state $|+\rangle$ 
is either a superposition of multiparticle states all composed
of even number of particles or all composed of odd number of particles. 
$|-\rangle$ is independent of the gauge field background and has an even
number of particles. If $|+\rangle$ is made up of states containing an odd number
of particles and the overlap is zero. Such a theory only makes sense if one couples
an even number of Majoran-Weyl fermions to the gauge field. If one couples two
Majorana-Weyl fermions, there is a potential for a fermion bilinear condenstate if
the background gauge field posses non-trivial mod(2) index.

\section{Monte Carlo evaluation of a fermion number violating observable}

The overlap formalism described in the previous section makes it possible to deal
with chiral gauge theories on the lattice. The overlap clearly passes all the
necessary tests needed of a correct formalism of the chiral determinant in a
fixed gauge field background. Physics under both perturbative and non-perturbative
gauge fields are correctly reproduced by the overlap in two and four 
dimensions~\cite{overlap,ictp}. To test the overlap formalism including the
dynamics of the gauge field is a non-trivial matter for two reasons: 
\begin{itemize}
\item Feasiblity to compute something non-perturbative in some model using the
overlap on the lattice using present day computers.
\item Existence of a chiral model where some non-perturbative results are known
by some other methods. Then one can see if the overlap reproduces these results.
\end{itemize}
The first reason is quite non-trivial due to the following points:
\begin{itemize}
\item Anomaly cancellation between different representations of Weyl fermions
occurs only in the continuum. On the lattice with a finite lattice spacing,
anomaly cancellation occurs only upto lattice spcaing effects. That is to say
the fermionic determinant on the lattice will have gauge violations which only vanish
as one takes the lattice spacing to zero. The existence of gauge violations on
the lattice implies that unphysical gauge degrees of freedom affect the dynamics
and if the overlap formalism has to reproduce a chiral gauge theory properly
on the lattice including the dynamics of gauge fields, then it has to be shown
that the unphysical gauge degrees of freedom do not affect the physics. 
Lattice gauge invariant theories have been shown to be robust under not too
large perturbations by gauge breaking terms~\cite{FNN}. In the overlap
formalism the gauge breaking appears only in the phase of the fermionic 
determinant and further it is small when the theory is anomaly free. 
The issue then is whether the gauge breaking is quantitatiely small for the
overlap formalism to result in the correct chiral gauge theory in the
continuum limit.
\item Difficulty in simulating theories with a complex action. The fermionic
determinant in a chiral gauge theory is usually complex and therefore 
standard Montecarlo simulations are not applicable.
\item Computation of the chiral determinant using the overlap involves
the diagonalization of $\ham^+$ which is a large matrix on any reasonable lattice.
\end{itemize}

With all the above issues in mind, a particular chiral U(1) model in two 
dimensions has proven to be a useful first testing ground for the overlap.
This model can be solved in the continuum following techniques similar to
that of the massless Schwinger model. There is a fermion number violating
process in this model. With a suitable choice of fermionic boundary conditions
on the torus the chiral determinant can be made real in the continuum. 
In the first subsection, the model is defined and the results from the
continuum computaton of the fermion number
violating process is presented. The computational details involved
on the lattice is then discussed. Results from the numerical simulation of this
model using the overlap on the lattice shows that the correct continuum
results are reproduced.

\subsection{11112 model on a continuum torus}

The 11112 model~\cite{chiral1,chiral2} is made up of a U(1) gauge field on a torus coupled
to four left handed Weyl fermions of charge $q=1$ and one left handed
Weyl fermion of charge $q=2$. 
The action is:
\begin{equation}
S={1\over 4e_0^2} \int d^2 x F^2_{\mu\nu}
-\sum_{f=1}^4\int d^2 x \bar\chi_f \sigma_\mu 
(\partial_\mu+iA_\mu )\chi_f
-\int d^2 x \bar\psi \sigma^*_\mu (\partial_\mu+2iA_\mu )\psi 
\end{equation}
where $\sigma_1=1$, $\sigma_2=i$ and $\mu =1,2$. The $U(1)$ gauge symmetry
is anomaly free by $2^2 = 1^2 +1^2 +1^2 +1^2 $. The boundary conditions are:
\begin{equation}
\begin{array}{rcl}
\chi_f(x + l_\mu \hat\mu) &=& e^{2\pi i b^f_\mu}\chi_f(x ) \\
\psi(x+ l_\mu \hat\mu) &=& \psi(x )\\
F_{\gamma\nu}(x+ l_\mu \hat\mu) &=& F_{\gamma\nu} (x)\\
\end{array}
\end{equation}
for $\mu =1,2$. $\hat\mu$ is a unit vector in the $\mu$ direction. The
$\bar\chi_f$ and $\bar\psi$ fields obey complex conjugate boundary conditions.
The $b_\mu^f$ are given by:
\begin{equation}
b^1_1=0;\ \ \ b^2_1=0;\ \ \ b^3_1={1\over 2};\ \ \ b^4_1={1\over 2};
\ \ \ \ \
b^1_2=0;\ \ \ b^2_2={1\over 2};\ \ \ b^3_2=0;\ \ \ b^4_2={1\over 2}.
\end{equation}
The chiral determinant is real and positive for the above choice of boundary
conditions~\cite{boundary}. The model can be solved in the continuum 
torus~\cite{chiralsolve}
following closely the technique for solving the massless Schwinger model
on the torus~\cite{Sachs}.

The massless sector consists of six left moving  Majorana
Weyl fermions forming a sextet under the global $SU(4)$ acting on the
$f$-index of the $\chi_f$'s. These particles are noninteracting. One can
choose interpolating fields for these particles which are neutral objects
local in the original fields:
\begin{equation}
\rho_{f_1 f_2} = - \rho_{f_2 f_1}= {{\pi^{3\over 2} e^{-\gamma}}\over
e_0} [\chi_{f_1} \chi_{f_2} \bar\psi - {1\over 2} \epsilon^{f_1 f_2 
f_3 f_4 } \bar\chi_{f_3 }
\bar\chi_{f_4 } \psi ] 
\end{equation}
The prefactor is chosen so that $\rho$ becomes a canonical field at large
distances. $\gamma$ is Euler's constant. 
The long distance behavior of the correlator is
\begin{equation}
\langle \rho_{f_1f_2}(0) \rho_{f_3f_4}(x) \rangle = \epsilon_{f_1f_2f_3f_4} 
{1\over 2\pi\sigma\cdot x}
\end{equation}
and there are two contributions to the correlator. This is due to the
non-trivial topology associated with U(1) gauge fields in 2D.
One contribution is from the
zero topological sector and is of the form
$$\langle \chi_{f_1}(0) \chi_{f_2}(0) \bar\psi(0) 
\bar\chi_{f_1}(x) \bar\chi_{f_2}(x) \psi(x) \rangle.$$
The second contribution is from the unit topological sector and is of the form
$$\langle \epsilon_{f_1f_2f_3f_4} \chi_{f_1}(0) \chi_{f_2}(0) \bar\psi(0) 
\chi_{f_3}(x) \chi_{f_4}(x) \bar\psi(x) \rangle.$$
The second contribution violates fermion number by two units. 
The exact low energy effective Lagrangian  of the model, written
in terms of the $\rho$-fields, is
\begin{equation}
{\cal L} ={1\over 2} \sum_{f_1 > f_2 } \rho_{f_1 f_2 } \sigma\cdot
\partial \rho_{f_1 f_2 }
\end{equation}
One of the terms in the effective
Lagrangian is a 't Hooft 
vertex, $V(x)$, which we choose to define as:
\begin{equation}
V(x)={{\pi^2}\over
{e_0^4}} \chi_1(x) \chi_2(x) \chi_3(x) \chi_4(x)
\bar\psi(x) (\sigma\cdot\partial )\bar\psi(x)
\end{equation}
This operator violates fermion number by two units and 
has a nonzero expectation value.
On the finite $t\times l$ lattice
\begin{equation} 
\langle V\rangle_{t\times l} = {64\pi \over (t m_\gamma)^4} 
\exp \Bigl[ -{4\pi\over  t m_\gamma}
\coth \left ( {1\over 2}lm_\gamma \right ) \Bigr] 
e^{4F( t m_\gamma)-8H( t m_\gamma , {t \over l} )}
\end{equation}
where $m_\gamma^2 = {{4e_0^2}\over \pi}$ and the functions $F(\xi )$ and 
$H(\xi , \tau )$ are defined below:
\begin{equation}
\begin{array}{rcl}
F(\xi ) & = & \sum_{n>0} \Bigl
[ {1\over n} - {1\over \sqrt{n^2+(\xi /2\pi)^2}}\Bigr] \\
H(\xi ,\tau) & = & \sum_{n>0}{1\over \sqrt{n^2+(\xi /2\pi)^2}}
{1\over e^{\tau\sqrt{(2\pi n)^2+\xi^2 }}-1} \\
\end{array}
\end{equation}
The infinite voulme limit of this quantity is 
\begin{equation}
\langle V\rangle_\infty = {e^{4\pi}\over 4\pi^3} \approx 0.081
\end{equation}
At $t=l={3\over m_\gamma}$
its value is 0.0389, substantially smaller than the value at infinite volume
and we will present results of simulations using the overlap formalism at this
finite volume. 

\subsection{Numerical results}

To perform the numerical simulation on the lattice~\cite{chiral2}
 and compute the fermion
number violating process, the expectation value $\langle V \rangle$ is
written on the lattice using the overlap as
\begin{equation}
\langle V \rangle =
{
{\int [dU] e^{S_g(U)} \langle-|V|+\wbrU \over \int [dU]e^{S_g(U)}}
\over
{\int [dU] e^{S_g(U)} \langle-|+\wbrU \over \int [dU]e^{S_g(U)}}
}
\end{equation}
Gauge fields are generated using the pure gauge action $S_g(U)$
and $\langle V \rangle$ is computed as a ratio of two observables
in the pure gauge theory, namely 
$\langle \langle-|V|+\wbrU\rangle$ and
$\langle\langle-|+\wbrU\rangle$. 
The pure gauge action is gauge invariant on the lattice but the
fermionic observable is not as mentioned in the beginning of this
section. Therefore one has to generate gauge fields configurations
according to the gauge invariant action to get points on the gauge
orbit 
and integrate the observable over many points on the gauge orbit
to account for the violation of the gauge symmetry. A finite physical
volume was maintained on the lattice by
setting $e_0=1.5{\sqrt{\pi}\over L}$ on a finite $L\times L$ lattice.

\begin{figure}
\centerline{\psfig{figure=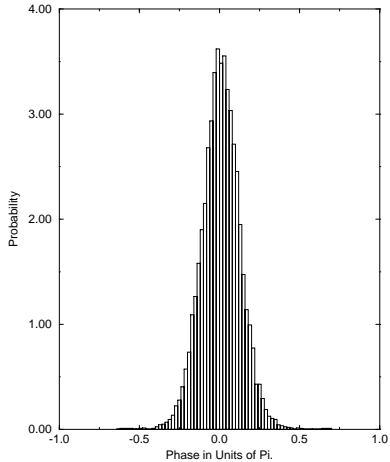,height=3in}}
\caption{
Phase distribution along a generic orbit carrying zero topological charge.
The horizontal axis is in units of $\pi$.  The phase is measured relative to
the Landau gauge phase and, within errors, the average cancels the Landau
phase leaving an almost real answer. The histogram contains 10,000
points.\label{fig:zero}}
\end{figure}

\begin{figure}
\centerline{\psfig{figure=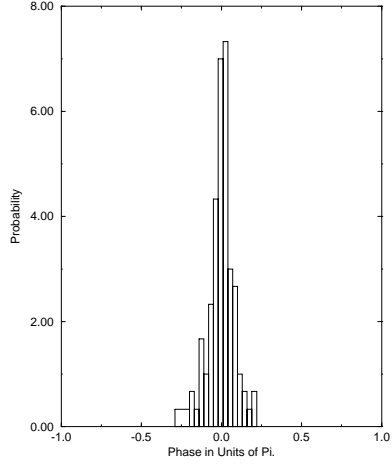,height=3in}}
\caption{
Phase distribution along a generic orbit carrying unit topological charge.
The horizontal axis is in units of $\pi$. The phase is measured relative to
the Landau gauge phase and, within errors, the average cancels the Landau
phase leaving an almost real answer. The histogram contains 100 points.
\label{fig:one}}
\end{figure}

As mentioned in the beginning of this section, the violation of
gauge invariance in the overlap is restricted to the phase.  
One can get a feel for the violation of gauge symmetry by plotting
the distribution of the phase of the overlap on a generic gauge orbit.
Such a distribution in zero topology and unit toplogy is shown 
in Fig.~\ref{fig:zero} and in Fig.~\ref{fig:one} respectively. Both figures
show that the distribution is well peaked around a central value
showing that the violations of the gauge symmetry are small.


\begin{figure}
\centerline{\psfig{figure=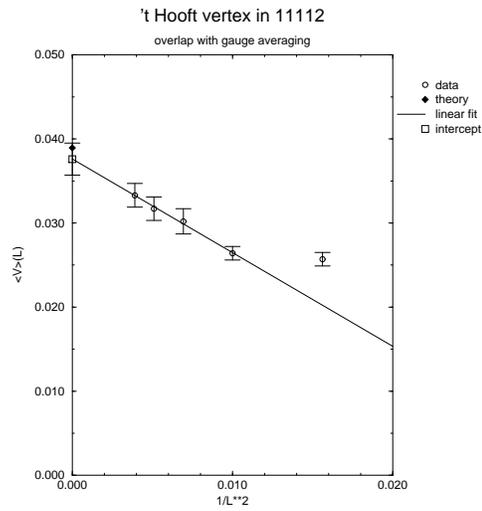,height=3in}}
\caption{
Data for $L=8,10,12,14,16$ versus $1/L^2 \propto a^2$, 
a linear
fit to the points $L\ge 10$, the continuum result (rhombus at $a=0$),
and our estimate for the continuum result from the data (square
with error bar at $a=0$)
\label{fig:vertex}}
\end{figure}

In Fig.~\ref{fig:vertex} the result for the computation of the 't Hooft vertex on
various lattices using gauge averaging on orbits is shown. The data fit well as
a function of $1/L^2$ and the continuum extrapolation matches well with the
number in the continuum. This gives a clear evidence that the overlap formalism
successfully reproduces the fermion number violating process in this model. 

In this model, there is a Thirring interaction that is marginal. If such a term
is generated, the continuum limit will not match with the
number obtained in the continuum without the Thirring term. The strength of
the Thirring term is regulator dependent and is
expected to depend on $m$ which is a regulator parameter
in the overlap formalism. It is conceivable that the Thirring term is small for
our particular choice of $m$ but it would be a miracle if it were exactly zero.
Evidence for the presence for a Thirring term will be deviations from the
$1/L^2$ behavior in Fig.~\ref{fig:vertex} for large enough $L$. This point was
investigated by going to $L=24$ and the data show some deviations in the
range of $L=20$ and $L=24$ but there is no clear evidence for a Thirring term 
yet~\cite{chiral2}.

\section {A numercial test of the continuum index theorem on the lattice}

The overlap formalism is capable of probing the topology of gauge fields on
the lattice as remarked in section 2.4. Given a gauge field configuration on
the lattice, there is an associated Hamiltonian $\ham^+$
ad given by Eq.~\ref{eq:lham} which is a $2K\times 2K$
matrix. If this matrix has $K-Q$ negative energy eigenstates then the index of
the chiral Dirac operator is $Q$. In particular, $Q$ units of fermions in
the fundamental representation of the gauge group are created by this gauge
field configuration. For a smooth configuration in the continuum, the Atiyah-Singer
index theorem~\cite{Atiyah} relates this to the topological charge of the gauge field.
Apriori, it is not obvious that this relation should hold on a finite lattice
away from the continuum since the configurations are not expected to be smooth.
It is possible to address this question in the context of pure gauge theory
if one has a measurement of the distribution
of topological charge and a measurement of the distrubution of the index 
for the same pure gauge action on the same lattice. A recent measurement of
the distribution of topological charge in pure SU(2) Yang-Mills theory using
the standard Wilson action has been performed 
using an improved colling method~\cite{cooling}.
On a $12^4$ lattice with $\beta=2.4$ this method gives a Gaussian distribution
with a variance of
$\langle Q^2\rangle = 3.9(5)$. Using the overlap, the distribution of the
index was also measured~\cite{index}
 on a $12^4$ lattice with $\beta=2.4$ using the standard
Wilson action and the variance of the distribution was $\langle Q^2\rangle = 3.3(4)$.
These two results show that relation between the index and the topological charge
holds quite well on a finite lattice away from the continuum.
In the following subsection, the details involved in measuring the index are
presented and the results are alo provided in some detail.

\subsection {Measurement of index and results}

The index of the chiral Dirac operator is directly related to the spectrum of
$\ham^+$ as given by Eq.~\ref{eq:lham}.
 In order to arrive at an efficient algorithm to measure the index
it is useful to study the spectral flow of 
\begin{equation}
\ham(\mu) = \gamma_5 (\dirac - \wilson + \mu) 
\label{eq:flow}
\end{equation}
where $\ham^+=\ham(m)$. 
In the chiral basis,
\begin{equation}
\ham(\mu) = \pmatrix{\mu-\wilson & \chiral \cr \chiral^\dagger & \wilson-\mu}
\end{equation}
and the condition for a zero eigenvalue of $\ham(\mu)$ is
\begin{equation}
\pmatrix{\mu-\wilson & \chiral \cr \chiral^\dagger & \wilson-\mu}
\pmatrix{ u\cr v\cr}=0\ \ \ \ \Leftarrow u^\dagger \wilson u + v^\dagger \wilson v = \mu
\end{equation}
implying that a solution can exist only for $\mu > 0$ since $\wilson$ is a positive
definite operator.
Therefore if $\ham^+$ has only $K-Q$ negative eigenvalues, then $\ham(\mu)$ should
have had $Q$ zero eigenvalues for certain values of $\mu$ between $0$ and $m$. That is
the spectral flow of $\ham(\mu)$ should show $Q$ levels crossing the axis from below
to above. This implies that one only needs to look at a few low lying eigenvalues
of $\ham(\mu)$ as a function of $\mu$ from $0$ to $m$ in order to measure the
index of the chiral Dirac operator. Several techniques are available to measure
a few low lying eigenvalues. One of them is the Lanczos method~\cite{Lanczos}.

\begin{figure}
\centerline{\psfig{figure=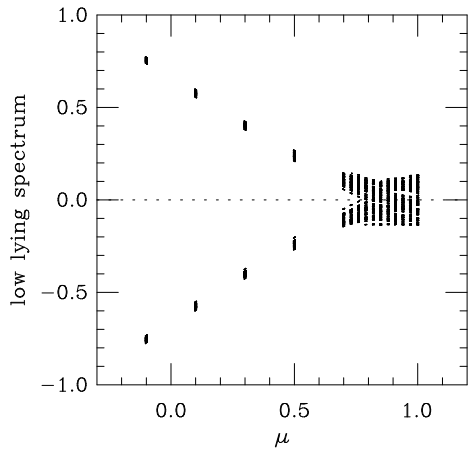,height=3in}}
\caption{
Five low lying positive and negative eigenvalues 
of $\ham(\mu)$ as a function of $\mu$ for five different gauge field
configuration. The configurations are from pure SU(2) gauge theory
at $\beta=2.4$ on a $12^4$ lattice.
\label{fig:flow1}}
\end{figure}
\begin{figure}
\centerline{\psfig{figure=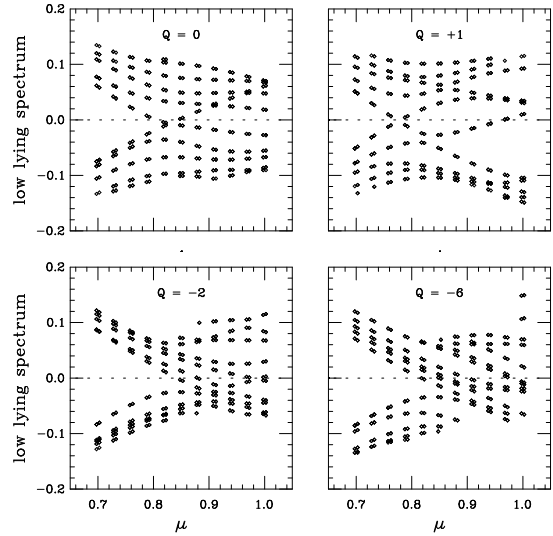,height=3in}}
\caption{
Five low lying positive and negative eigenvalues 
of $\ham(\mu)$ as a function of $\mu$ for four gauge field
configurations with different values of $Q$. The configurations are
from pure SU(2) gauge theory at $\beta=2.4$ on a $12^4$ lattice .  
\label{fig:flow2}}
\end{figure}
\begin{figure}
\centerline{\psfig{figure=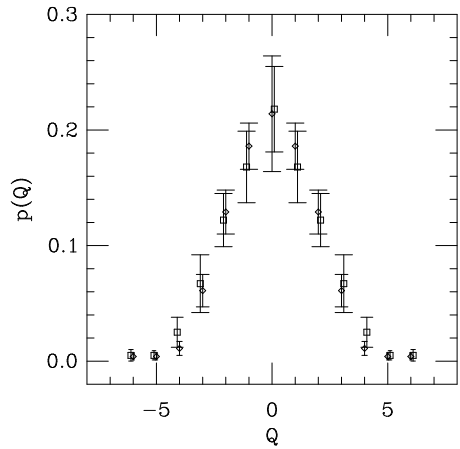,height=3in}}
\caption{
Comparison of the
probability distribution of the index 
(diamonds)
with the probability distribution of the topological charge(squares).
The squares have been slightly shifted laterally
for visual purposes.
\label{fig:index}}
\end{figure}
\begin{table}[t]
\caption{Comparison of the
probability distribution of the index 
with the probability distribution of the topological charge.
\label{tab:index}}
\vspace{0.4cm}
\begin{center}
\begin{tabular}{|c|c|c|}
\hline
 & &  \\
$Q$&$p(Q)$(Index)&p(Q)(Topology) \\ 
 & & \\
\hline
 & & \\
0&0.214(50)&0.218(37) \\
$\pm$1&0.186(20)&0.168(31) \\
$\pm$2&0.129(19)&0.122(23) \\
$\pm$3&0.061(14)&0.067(25) \\
$\pm$4&0.011(6)&0.025(13) \\
$\pm$5&0.004(3)&0.005(4) \\
$\pm$6&0.004(3)&0.005(5) \\
 & &  \\ \hline
 & &  \\
$\langle Q^2\rangle$&3.3(4)&3.9(5) \\
 & &  \\
\hline
\end{tabular}
\end{center}
\end{table}

In Fig~\ref{fig:flow1} a few low lying eigenvalues of the $\ham(\mu)$ as a function
of $\mu$ is shown for a few configurations in the gauge field ensemble. 
The spetrum has a gap for $\mu<0.7$ and the gap closes
around this value. Fig.~\ref{fig:flow2} provides a closer look at the flow of
four different configurations around the region in $\mu$ where the gap closes.
From this one can read the index associated with the configuration. Clearly
all crossings do not happen at some fixed value of $\mu$. The finite spread in
$\mu$ is a consequence of finite lattice spacing. For smooth instantons embedded on
a finite lattice, one can show that smaller instantons cross later in
 $\mu$~\cite{overlap,prep}. The region of $\mu$ where the crossing happen
will get closer to $\mu=0$ as one approaches the continuum limit and the
spread in $\mu$ will also shrink~\cite{index}. On a finite lattice, one can
use the spread to infer some information of the shape distribution of topological
objects on the lattice. 

In Table~\ref{tab:index} the distribution for the topological charge using
improved cooling~\cite{cooling} is listed along with the distribution for
the index obtained using the overlap~\cite{index}. The two colums are a
result of measurements on a different set of independent configurations.
The table is plotted in Fig.~\ref{fig:index}. The close matching of
the two distributions indicate that the cpnnection between the index and the
topological chrge remains valid on the lattice in a probabilistic sense.

\section{Conclusions and future directions}

Montecarlo measurement of a fermion number violating process in a two dimensional
chiral model and a test of the continuum index theorem on the lattice in a four
dimensional gauge theory has shown that the overlap formalism is a valid 
proposal to deal with chiral fermions in a non-perturbative manner. Future work
using the overlap has to be planned by keeping the various points mentioned
in the beginning of section 3. Problems in two dimensions such as Majorana-Weyl
fermions coupled to non-abelian gauge fields could be done without worrying about
new stochastic algorithms for the fermions. 

To make the overlap formalism into
a viable technique in four dimensions one has to think of new algorithms to deal
with the overlap. This has to be done in several steps. The first step has already
been taken. This is the measurment of the index in pure gauge theory. 
Fermion dynamics do not play a part and one can use the well developed
techniques to deal with gauge dynamics. The measurement of the index is a computation
of a fermionic obsevable but it is one step simpler than the computation of a
fermionic correlator. Efficient techniques to deal with low lying eigenvalues of
$\ham^+$ made it possible to measure the index. It would be useful to obtain
the continuum distrubution of the index in four dimensional
pure gauge theories using the overlap.
Without much more effort, one
could also compute the associated eigenvectors. These eigenvectors will carry
information about localized objects. It is plausible that these low lying
eigenvectors carry most of the physics information. This would be the case for
instance if physics is driven by instantons~\cite{Shuryak}. Therefore the second
step in four dimensions would be to deal with quenched QCD and use an approximate
form of the overlap (keeping a few low lying eigenstates of $\ham^+$) to measure
fermionic correlations. Based on the progress made in the second step one could
push further to deal with massless QCD using the overlap. In the context of
QCD, the operator does have an eigenvalue problem and it is possible to
reduce the overlap formula to the computation of the determinant of a finite
matrix~\cite{finite}.
Only after this could
one try to deal with chiral gauge theories in four dimensions since one then has
to deal with complex action. This is a long path but the ability to deal with
chiral fermions in principle makes it possible to start the hike. 

\vspace*{-2pt}
\section*{Acknowledgments}
The author is deeply indebted to Herbert Neuberger for a very fruitful collaboration
which began over five years ago. Along the way, the author has benifited from
useful discussions with 
Luis Alvarez-Gaume, Peter Arnold, Khalil Bitar, Philippe de Forcrand,
Robert Edwards, Anna Hasenfratz, Urs Heller, Patrick Huet, David
Kaplan, Tony Kennedy, Yoshio Kikukawa, Chris Korthals-Altes, Tamas
Kovacs, Martin L\"uscher, Aneesh Manohar, Margarita Garc\'ia P\'erez,
Michael Peskin, Seif Randjbar-Daemi, Adam Schwimmer, Steve Sharpe, Bob Singleton,
Andrei Slavnov, Jan Smit, Nucu Stamatescu, Massimo Testa, Pavlos
Vranas, Frank Wilczek, and Larry Yaffe.
The authors wishes to thank all the local organizers of the AIJIC 97 conference
for their time and effort in organizing a pleasant and useful meeting.

\eject

\end{document}